\documentclass[aps,prl,reprint,amsmath,amssymb,groupedaddress]{revtex4-2}
\usepackage{graphicx}
\usepackage[colorlinks,
linkcolor=blue,
anchorcolor=blue,
citecolor=blue,urlcolor=blue]{hyperref}
\begin{document}

	\title{Revisiting Multi-Wave Resonance in Classical Lattices: Quasi-Resonances, Not Exact Resonance, Govern Energy Redistribution}

\author{Wei Lin}
	\author{Yong Zhang}
	\author{Hong Zhao}
	\email{E-mail: zhaoh@xmu.edu.cn}
\affiliation{$^1$Department of Physics, Xiamen University, Xiamen 361005, Fujian, China}
\date{\today}

\begin{abstract}
The multi-wave exact resonance condition is a fundamental principle for understanding energy transfer in condensed matter systems, yet the dynamical evolution of waves satisfying this condition remains unexplored. Here, we reveal that the multi-wave resonant kinetic equations possess distinctive symmetry properties that preferentially induce energy equalization between counter-propagating waves of identical frequency. This initial equalization disrupts the exact resonance condition, rendering it dynamically invalid. We further demonstrate that nonlinearity-mediated multi-wave quasi-resonances—not exact resonances—govern energy transfer and drive the system toward thermalization. Crucially, the strength of exact resonances decays with increasing system size, while quasi-resonance strength grows. Moreover, exact resonance strength remains independent of nonlinearity, whereas quasi-resonance strength diminishes with reduced nonlinearity. These observations provide additional evidence supporting the aforementioned conclusion while elucidating the size-dependent thermalization characteristics in lattice systems.
\end{abstract}

\maketitle

	\maketitle
	
	\section{Introduction}
The $n$-phonon resonance conditions~\cite{Peierls:1929,Dalitz:1997}:
	\begin{equation}\label{eq:kw}
		k_1 \pm k_2 \dots \pm k_n \ \text{mod} \ N = 0 \quad \text{and} \quad \omega_{k_1} \pm \omega_{k_2} \dots \pm \omega_{k_n} = 0
	\end{equation}
serve as fundamental tools for studying energy diffusion and transport in condensed matter systems~\cite{Chaplain:2020,Qian:2021,Ravichandran:2021,Li:2022,Ravichandran:2019}, where $k_i$ and $\omega_{k_i}$ denote the wavevector and frequency of the $i$-th phonon, respectively. Their classical counterpart, the multi-wave exact resonance, shares an identical mathematical form with Eq.~(\ref{eq:kw}) but replaces "phonons" with "normal modes" or "waves" in classical lattices. While conventional theory assumes simultaneous energy exchange among modes satisfying these conditions, the actual dynamical evolution of resonant waves has not yet been directly revealed through either numerical simulations or experimental observations.

Here we demonstrate that exact resonance is not the primary mechanism for energy transfer or diffusion in classical lattice systems, since the time evolution breaks the resonance conditions. This breakthrough stems from identifying symmetric structures in the governing kinetic equations that partition resonance sets into independent mode subsets. These subsets evolve toward separate steady states, invalidating the kinetic equations and halting exact resonance. The preferential energy equalization between counter-propagating wave pairs of identical frequencies (hereafter referred to as "pairwise equalization") constitutes the primary mechanism for terminating exact resonances. We further confirm that it is the multi-wave quasi-resonances induced by nonlinear interactions, constitutes the dominant mechanism driving energy redistribution and facilitating thermalization toward equilibrium in classical lattice systems. In addition to the discovered symmetry in the kinetic equations, another key finding substantiates this conclusion: the exact multi-wave resonance strength decays with reduced system size, whereas the quasi-resonance intensity increases with system size. This implies that exact resonance effects can be neglected in sufficiently large systems. In addition, we report that the quasi-resonance strength decreases with diminishing nonlinear interaction intensity, while the exact resonance remains size-independent. These findings demonstrate that both exact and quasi-resonances remain effective within characteristic system sizes, providing critical insight into the size-dependent energy relaxation dynamics

The perspective that quasi-resonance plays a pivotal role ultimately synthesizes the collective wisdom gained from over half a century of research on the celebrated Fermi-Pasta-Ulam-Tsingou (FPUT) problem~\cite{Fermi:1955,Dauxois:2008,Lepri:1998,Poggi:1995,Luca:1995,Luca:1999,Parisi:1997,Casetti:1997,Cretegny:1998,Benettin:2011,Flach:2010,Laptyeva:2014}, with particular relevance to contemporary advances in wave turbulence theory. The FPUT problem has stimulated advances in integrability ~\cite{Zakharov:1991} and soliton physics ~\cite{Zabusky:1965}, yet the original question of how lattice modes interact to achieve thermalization was only systematically clarified in the past decade through wave turbulence theory ~\cite{Onorato:2015,Lvov:2018,Pistone:2018,Fu:2019,Fu:2019R,Pistone:2019,Wang:2020,Onorato:2023,Wang:2024,Lin:2025}.  Originating from Peierls’ 1929 derivation of kinetic equations (Eq.~\ref{eq:kw}) ~\cite{Peierls:1929,Dalitz:1997}, wave turbulence theory finds broad applications—from oceanography ~\cite{Aubourg:2017,Zakharov:2019} to Bose-Einstein condensates ~\cite{Connaughton:2005,Sun:2012,Zhu:2023}. Current understanding, based on this theory, establishes that thermalization in the thermodynamic limit under infinitesimal non-integrable perturbations depends critically on eigenmode types: lattices with extended modes obey universal time-scaling laws ~\cite{Pistone:2018,Fu:2019,Fu:2019R,Pistone:2019,Wang:2020,Wang:2024}, whereas fully localized systems exhibit diverging thermalization times, leading to thermal glass states ~\cite{Lin:2025}. These studies necessitate distinguishing between exact resonance and quasi-resonance effects. Crucially, quasi-resonance—transient multi-wave interactions enabled by nonlinear frequency broadening ~\cite{Zaslavskii:1998,Holloway:1980,Pan:2017,Pushkarev:1999,Connaughton:2001,Kartashova:2007,Lvov:2010}—plays an indispensable role in lattice thermalization ~\cite{Pistone:2018,Pistone:2019, Wang:2020,Lin:2025,Wang:2024}, with its impact sometimes seeming to surpass exact resonance ~\cite{Kartashova:2007,Lvov:2010,Nazarenko:2011,Lin:2025}. The present work makes three new fundamental contributions: (i) unambiguously demonstrating the predominance of quasi-resonances in thermalization processes, (ii) revealing the pivotal role of pairwise equalization in driving systems toward equilibrium, and (iii) demonstrating the existence of a finite threshold for nonlinear perturbations that enables thermalization in finite-sized FPUT-type systems—contrary to exact resonance analyses predicting small system can always been thermalized via 6-wave resonances ~\cite{Onorato:2015,Lvov:2018}. 

\section{Our Models}
In this paper, we focus on demonstrating the symmetry recovery mechanism using a specific lattice model, while also establishing its universality across a range of lattice systems through additional models.  The specific model we employed is the 6th-order nonlinear potential FPUT model, abbreviated as the FPUT-6 lattice, which Hamiltonian is given by
	\begin{eqnarray}\label{eq:H6}
		H = \sum_{j} \frac{p_j^2}{2}  + \frac{1}{2}(q_j - q_{j+1})^2 + \frac{1}{6}b (q_j - q_{j+1})^6,
	\end{eqnarray}
where $p_j$ and $q_j$ represent the momentum and displacement of the $j$-th particle, respectively. The constant $b$ modulates the strength of the nonlinear perturbation. We rescale $q_j$ by the energy density $\epsilon$, defining $q_j = \tilde{q}_j \epsilon^{1/2}$, and rewrite the Hamiltonian as ${\tilde H}=H/\epsilon=H_0(\tilde{q}_j, \tilde{p}_j) + \sum_j \frac{1}{6}g \tilde{q}_j^{6}$, where $g = b\epsilon^2$ represents the nonlinearity strength. 

As previously established ~\cite{Onorato:2015,Lvov:2018}, a connected network of 6-wave exact resonances always exists in generic FPUT lattices. These exact resonances are independent of nonlinear interaction strength, implying that general finite lattice systems can always approach equilibrium and achieve thermalization through at least 6-wave resonances. To provide definitive evidence for the failure of exact resonances in dynamical processes and consequently challenge the prevailing view, we have developed this specialized model.
Both 6-wave exact resonance and 6-wave quasi-resonance exhibit similar time relaxation scaling laws. Therefore, if we only observe the 6-wave resonance time scaling in a model, we cannot determine which type of resonance is responsible for it. However, using this model, we can demonstrate that as the nonlinearity strength decreases, the two resonance effects appear successively, and which allows us to track the dynamical breakdown of exact-resonance-derived scaling laws. The commonly used template models, namely the FPUT-$\alpha$ and FPUT-$\beta$ models, are not ideal for this purpose, as lower-order resonances interfere with the 6-wave resonance effects. However, We validate through these canonical models that all lattice systems ultimately exhibit exact resonance failure.

	\section{Kinetic Equations and 6-wave Exact Resonance Solutions}
	
	Due to the 6th-order nonlinear interaction term, the fundamental kinetic equations are 6-wave resonance equations, which can be written as
	\begin{eqnarray}\label{eq:H6keq}
		\begin{split}
			\dot{D}_{k_1}= & -2g^2 \pi \sum_{k_2,...,k_6} A^2_{k_1k_2k_3k_4k_5k_6} (C^{k_2k_3k_4k_5k_6}_{k_1}\omega^{k_2k_3k_4k_5k_6}_{k_1}    \\
			&  + 5C_{k_1k_2}^{k_3k_4k_5k_6}\omega_{k_1k_2}^{k_3k_4k_5k_6}  +  10C^{k_4k_5k_6}_{k_1k_2k_3}\omega_{k_1k_2k_3}^{k_4k_5k_6}   \\
			&  + 10C_{k_1k_2k_3k_4}^{k_5k_6}\omega_{k_1k_2k_3k_4}^{k_5k_6} + 5C_{k_1k_2k_3k_4k_5}^{k_6}\omega^{k_6}_{k_1k_2k_3k_4k_5}\\
			& +C_{k_1k_2k_3k_4k_5k_6}\omega_{k_1k_2k_3k_4k_5k_6})\\
			= & \eta_{1} - \gamma_{1}D_{k_1}.
		\end{split}
	\end{eqnarray}
	Here, $D_{k_1} = \langle a_{k_1} a^*_{k_1} \rangle$ with $a_{k_1} = \frac{1}{\sqrt{2\omega_{k_1}}}\left(\omega_{k_1} Q_{k_1} + i P_{k_1}\right)$; $Q_{k_1}$ and $P_{k_1}$ are the amplitude and conjugate momentum of the $k_1$-th normal mode. $A_{k_1 k_2 k_3 k_4 k_5 k_6} = \pm \frac{\sqrt{\omega_{k_1} \omega_{k_2} \omega_{k_3} \omega_{k_4} \omega_{k_5} \omega_{k_6}}}{64 N^2} \delta(k_1 \pm k_2 \pm k_3 \pm k_4 \pm k_5 \pm k_6),
	$
	while the interaction coefficient $
	C^{k_4 k_5 k_6}_{k_1 k_2 k_3} = 120 D_{k_1} D_{k_2} D_{k_3} D_{k_4} D_{k_5}D_{k_6} $ $ \left( \frac{1}{D_{k_4}} + \frac{1}{D_{k_5}} + \frac{1}{D_{k_6}} - \frac{1}{D_{k_1}} -\frac{1}{D_{k_2}} - \frac{1}{D_{k_3}} \right),
	$
	with the frequency resonance condition:
	$
	\omega_{k_1 k_2 k_3}^{k_4 k_5 k_6} = \delta (\omega_{k_4} + \omega_{k_5} + \omega_{k_6} - \omega_{k_1} - \omega_{k_2} - \omega_{k_3}).
	$
	Other terms within the brackets can be written similarly. The energy of the $k_1$-th mode is given by $E_{k_1} = D_{k_1} \omega_{k_1}$. 
	
	A detailed derivation is provided in Methods. The stochastic phase and stochastic amplitude assumptions are applied in the derivation, commonly used for similar purposes~\cite{Nazarenko:2011}. Consistent with prior analogous research ~\cite{Wang:2020,Onorato:2023}, it can be found that in Eq.~(\ref{eq:H6keq}), both $\eta_{1}$ and $\gamma_{1}$ are independent of $D_{k_1}$ and, in this model, are proportional to $g^2$. This leads to a scaling of the relaxation time $T_c \propto 1/\gamma_{1} \sim g^{-2}$ for mode $k_1$, provided that $\gamma_{1}$ is a non-vanishing constant. 
	
	Satisfying the multi-wave exact resonance conditions~(\ref{eq:kw}) with $n=6$ guarantees the non-vanishing of $\gamma_{1}$, because this leads to non-zero coefficients $A_{k_1 k_2 k_3 k_4 k_5 k_6}$ and non-zero terms inside the parentheses on the right-hand side of Eq.~(\ref{eq:H6keq}). The six terms inside the parentheses correspond to the processes 1-5, 2-4, 3-3, 4-2, 5-1, and 6-0 in 6-wave resonances. Due to the dispersion relation constraints of these lattices, there are no exact resonance solutions for the 1-5, 5-1, and 6-0 processes. For systems where $N$ isn't divisible by 3, there are no exact resonance solutions for the 4-2 and 2-4 processes either. The only solutions are the 
	so-called symmetric 3-3 resonance $(k_1, k_2, k_3, -k_1, -k_2, -k_3)$ with $k_1 + k_2 + k_3 = lN/2$, and quasi-symmetric 3-3 resonances $(k_1, k_2, k_3, -k_1, -k_2, k_3)$ with $k_1 + k_2 = lN/2$, where $l$ is an integer ~\cite{Onorato:2015}. For systems where $N$ is divisible by 3, in addition to the symmetry and quasi-symmetry solutions, there also exist so-called non-pairing solutions~\cite{Bustamante:2019} (The generality of the conclusions is not affected by excluding these solutions, see Supplementary Section 1). These approaches can be generalized to determine the resonance solutions for general $n$-wave interactions. This knowledge represents a significant contribution of the wave turbulence approach to the understanding of multi-wave resonance processes.
	
\section{Symmetry-Induced Constraints in Kinetic Equations and the Collapse of Exact Resonances}

In this section, we fix the system size at N=32 to demonstrate the evolutionary dynamics of exact resonances. we can then search for all exact resonant 6-wave solutions according to the resonance conditions. $(k_1, k_2, k_3, k_4, k_5, k_6) = (6,12,14,-6,-12,-14)$ and $(k_1, k_2, k_3, k_4, k_5, k_6) = (1,6,9,-1,-6,-9)$ are two examples of 3-3 symmetry solutions. Based on the traditional understanding of multi-wave resonance, it would be expected that modes within each set exchange energy among themselves, and these two sets would exchange energy through the common mode pair $(6,-6)$. However, we will demonstrate that it is not as simple as that. 
	
In Fig.~1 (a), we illustrate the time evolution of $D_k$ for the first set of modes at $g = 10^{-2}$ following the kinetic equations. All of the 3-3 symmetry solutions are involved in the evolution of the kinetic equations. Simulation details provided in Methods. As seen in the figure, after evolving for some time, $D_k$ and $D_{-k}$ converge to their average values and cease to evolve further. We refer to this phenomenon as symmetry recovery, meaning the energies of the paired normal modes regain symmetry.
	
The 3-3 quasi-symmetry solutions include examples such as $(6,10,2,-6,-10,2)$ and $(2,14,6,-2,-14,6)$. Considering all quasi-symmetry solutions, the evolution of $D_k$ is shown in Fig.~1 (b). We observe that while energy in the pairs of $-k$ and $k$ does not reach the same value, the evolution reaches a stable state. When simultaneously evolving all symmetry and quasi-symmetry solutions, we observe complete symmetry recovery across all modes (Fig.~1(c)), albeit with a slightly modified dynamical pathway compared to Fig.~1(a).
	
	\begin{figure}[htb] 
		\centering
		\includegraphics{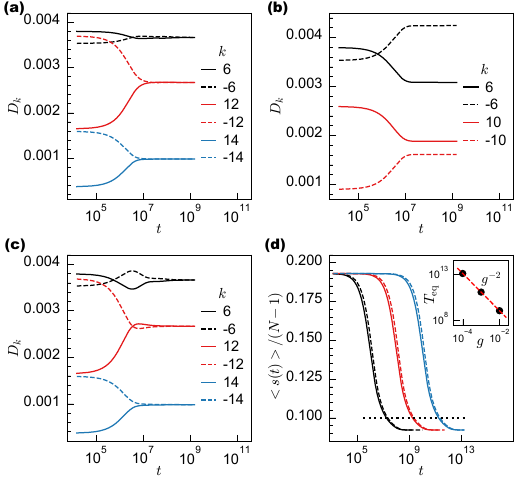}
		\caption{Exact resonance dynamics under symmetry constraints. (a)--(c) show the time evolution of $D_k$ under the kinetic equations for the 3-3 symmetry solutions, the 3-3 quasi-symmetry solutions, and the full set of solutions, for $N = 32$ at $g = 10^{-2}$. (d) displays the indicator entropy at different values of $g$. Dashed lines represent results restricted to the 3-3 symmetry solutions, while solid lines correspond to all solutions. From left to right, $g$ decreases with values $g = 10^{-2}, 10^{-3}, 10^{-4}$. The inset shows the relationship between the relaxation time and $g$, with the dashed line indicating the scaling law $T_c \propto g^{-2}$ as a reference.}
       \label{fig:jq}
	\end{figure}
	
Although energy is exchanged mainly in pairs, the kinetic equations~(\ref{eq:H6keq}) remains valid before exact recovery, as $\gamma_{1}$ is non-zero in this case. This means the system's evolution follows the scaling law $T_c\sim g^{-2}$ during this stage. In Fig.~\ref{fig:jq}(d), we present the time evolution of the indicator entropy $\langle s(t) \rangle/(N-1)$ (defined in Methods) at several levels of $g$. The solid curves represent numerical results from evolving the kinetic equations that include all 3-3 solutions. The entropy $\langle s(t) \rangle/(N-1)$ can be used to measure the degree of energy equipartition, approaching 0 as the system approaches energy equipartition.

We see that, rather than evolving towards zero, the entropy converges to $0.09$, independent of $g$. This result indicates that the resonance process has ceased. However, if we use a threshold of $\langle s(t) \rangle/(N-1) = 0.1$, which is slightly higher than the saturation value, to calculate the relaxation time, we find $T_c \propto g^{-2}$ (see the inset in Fig.~1(d) for confirmation). Therefore, before the resonance process stops, the system's energy relaxation follows the scaling law of 6-wave resonance. 
	
It can be prove that the maximum value of $\langle s(t) \rangle/(N-1) \sim 0.19$ originates from the initial random sampling of $E_k$ from a uniform distribution. It can further show that $\langle s(t)\rangle/(N-1)$ asymptotically approaches $0.091$ when the evolution terminates at $(E_k + E_{-k})/2$ for each symmetry pair (Supplementary Section 2). This indicates that the initial energy relaxation primarily occurs through symmetry recovery—each $k$ and $-k$ mode pair first achieves energy equivalence (thereby forming standing waves). This process generates sufficiently low $\langle s(t) \rangle$ values, and if the threshold is set too high, the symmetry recovery phenomenon will not be detectable. In addition, the dashed lines in Fig.~1(d) represent the evolution of only the 3-3 symmetry solutions. The difference from that by involving all 3-3 solutions is minimal, indicating that the contribution of the quasi-symmetry solutions is relatively small.
	
We now explain why the 6-wave exact resonances exhibit such behavior. For the 3-3 symmetry solutions $(k_1,k_2,k_3,-k_1,-k_2,-k_3)$, Eq.~(\ref{eq:H6keq}) can be simplified to
	\begin{eqnarray}\label{eq:6keq-jq}
		\begin{split}
			\dot{D}_{k_1} =& -\frac{75 g^2 \pi} {128 N^4 } \sum_{k_2,k_3}  \omega^2_{k_1} \omega^2_{k_2} \omega^2_{k_3} D_{k_1} D_{k_2} D_{k_3} D_{-k_1} D_{-k_2}D_{-k_3}\\
			&\times \left(\frac{1}{D_{-k_1}} + \frac{1}{D_{-k_2}}+ \frac{1}{D_{-k_3}} -\frac{1}{D_{k_1}}-\frac{1}{D_{k_2}} -\frac{1}{D_{k_3}}\right).
		\end{split}
	\end{eqnarray}
It is obvious that the counter-propagating wave set $(-k_1,-k_2,-k_3,k_1,k_2,k_3)$ is also a solution to the 3-3 symmetry if $(k_1,k_2,k_3,-k_1,-k_2,-k_3)$ is. Then we have $\frac{d}{dt} \left( D_{k_1} + D_{-k_1} \right) = 0$, and therefore $D_{k_1} + D_{-k_1} = \text{constant}$ or $\dot{D}_{k_1} = -\dot{D}_{-k_1}$. These findings demonstrate that counter-propagating modes in the two resonance sets form isolated energy-exchange pairs, exhibiting fully decoupled dynamics from other mode pairs. As a result, the evolution must stop when $D_{k_1} = D_{-k_1}=\left( D_{k_1} + D_{-k_1} \right) / 2$, which explains the results shown in Fig.~\ref{fig:jq}(a).
	
	The kinetic equations for the 3-3 quasi-symmetry solution can be simplified to
	\begin{eqnarray}\label{eq:6keq-jq2}
		\begin{split}
			\dot{D}_{k_1} &= \frac{75g^2 \pi}{128 N^4}     \omega^2_{k_1}\omega^2_{k_2} D_{k_1} D_{-k_1} D_{k_2} D_{-k_2}\\ 
			&\times \left( \frac{1}{D_{k_1}}+\frac{1}{D_{k_2}} - \frac{1}{D_{-k_1}} - \frac{1}{D_{-k_2}}\right) \sum_{k_3} \omega^2_{k_3} D_{k_3}^2.
		\end{split}
	\end{eqnarray}
	Here, $k_2 = \frac{N}{2} - k_1$ for $k_1 > 0$, and $k_2 = -\frac{N}{2} - k_1$ for $k_1 < 0$. These solutions also exhibit special symmetry. For example, the sets $(6, 10, k_3, -6, -10, k_3)$, $(-6, -10, k_3, 6, 10, k_3)$, $(10, 6, k_3, -10, -6, k_3)$, and $(-10, -6, k_3, 10, 6, k_3)$ are all 3-3 quasi-symmetry solutions. Additionally, there are no restrictions on $k_3$, so the summation over $k_3$ is the same for all $\dot{D}_{k_1}$. This symmetry property leads to $\frac{d}{dt} \left( D_{k_1}+D_{k_2} -D_{-k_1} -D_{-k_2} \right) =0$. Consequently, energy exchange occurs between pairs of ($E_{k_1}, E_{k_2}$) and ($E_{-k_1}, E_{-k_2}$), and the evolution stops when $\frac{1}{D_{k_1}}+\frac{1}{D_{k_2}}=\frac{1}{D_{-k_1}} + \frac{1}{D_{-k_2}}$. This process does not require the energy of $-k$ and $k$ modes to reach the same value, explaining the results in Fig.~\ref{fig:jq}(b).
	
	However, due to the existence of 3-3 symmetry solutions, a counter-propagating modes $k$ and $-k$ will ultimately converge to the same energy as a result of the joint evolution of Eqs.~(\ref{eq:6keq-jq2}) and~(\ref{eq:6keq-jq}); otherwise, the evolution will not cease. This fact explains Fig.~\ref{fig:jq}(c). In addition, 3-3 symmetry solutions have more freedom in constructing $k_1, k_2, k_3$ under the condition $k_1 + k_2 + k_3 = lN/2$, compared to the 3-3 quasi-symmetry  solutions where $k_1$ and $k_2$ are chosen under the condition $k_1 + k_2 = lN/2$, and $k_3$ does not participate in the energy exchange. This explains why 3-3 quasi-symmetry solutions contribute little to energy relaxation, as seen in Fig.~\ref{fig:jq}(d), and the pairwise equalization between counter-propagating modes $(k, -k)$ governs the kinetic equation's evolution. 
	
	Note that although $k_3$ does not participate in the energy exchange, the quasi-symmetry solutions cannot be regarded as 4-wave resonances. The mode $k_3$ appears in the kinetic equations, and if its amplitude were zero, the 6-wave kinetic equations would fail. Similarly, in symmetry 3-3 resonance sets, although the two modes in a pair share the same frequencies and are thus degenerate, symmetry recovery differs from the usual Chirikov resonance~\cite{Izrailev:1966,Chirikov:1979}, as it requires two other pairs of modes to satisfy the 6-wave resonance conditions.
	
	\section{Quasi-Resonance vs Exact Resonance}
	
	Nonlinear interactions cause frequency broadening and leads to quasi-resonance conditions[]:
    
	\begin{equation}\label{eq:zhun}
		\begin{split}
			&\lvert \omega_{k_1} \pm \omega_{k_2} \pm \omega_{k_3} \pm \omega_{k_4} \pm \omega_{k_5} \pm \omega_{k_6} \rvert < \Omega \\
			&\& \ \ A_{k_1 k_2 k_3 k_4 k_5 k_6} \neq 0.
		\end{split}
	\end{equation}
	
	As the quasi-resonance parameter $\Omega \rightarrow 0$, these conditions converge to the exact resonance conditions. Unlike exact resonances, quasi-resonances depend on the strength of nonlinear interactions. We define the connectivity strength of mode $k_1$ as~\cite{Lin:2025}:
	\begin{equation}\label{eq:p6}
		\begin{split}
			p_6(k_1) =&\sum_{k_2 \dots k_6} |A_{k_1k_2k_3k_4k_5k_6}| \\
			=&\sum_{k_2 \dots k_6} \frac{\sqrt{\omega_{k_1} \omega_{k_2} \omega_{k_3} \omega_{k_4} \omega_{k_5} \omega_{k_6}}}{64 N^2} \delta_{k_1k_2k_3}^{k_4k_5k_6},
		\end{split}
	\end{equation}
to quantify the strength of resonances. Here, the summation runs over all modes satisfying the quasi-resonance conditions for mode $k_1$. The value $p_6(k_1)$ measures the ability of the $k_1$-th mode to diffuse energy through 6-wave resonances. Physically speaking, $\Omega$ must be a monotonically varying function of $g$. In this model, $\Omega \sim g^2$ (Supplementary Section 3). For $\Omega = 0$, $p_6(k_1)$ can also be calculated, providing insight into the strength of exact resonances.
	\begin{figure}[tb] 
		\centering
		\includegraphics{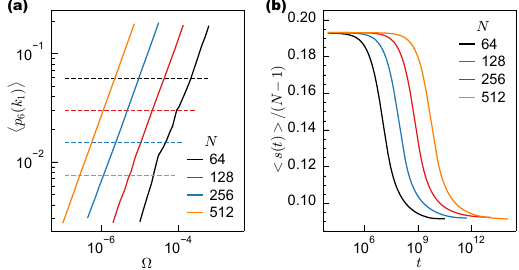}
		\caption{Quasi-resonance vs. exact resonance. (a) shows the dependence of average connectivity on $\Omega$. Dashed lines represent the results for 6-wave exact resonances, while solid lines represent 6-wave quasi-resonances. (b) shows the indicator entropy as a function of time for different system sizes at $g = 10^{-2}$, calculated using the kinetic equations including all allowed 3-3 modes.}
        \label{fig:zhun} 
	\end{figure}
	
Figure~\ref{fig:zhun}(a) displays the $\Omega$-dependence (equivalently, $g$-dependence) of the average connectivity strength for 6-wave quasi-resonances (excluding exact resonance contributions) and exact resonances in FPUT-6 lattices of varying sizes. We see that at fixed $\Omega$, quasi-resonance connectivity strengthens with increasing system size. On the other hand, the exact resonance connectivity strength exhibits an inverse size dependence, implying that exact-resonance-induced relaxation dynamics evolve progressively slower in larger systems and will vanish in the thermodynamic limit. This prediction is quantitatively verified by the kinetic equations simulations presented in Fig.~\ref{fig:zhun}(b), where the relaxation time induced by exact resonance increases with system size. The fact that $\langle s(t) \rangle/(N-1)$ converges to the critical value of $0.09$ indicates that even in thermodynamic limit systems, the contribution of exact resonances to energy diffusion only achieves direct energy equilibration between counter-propagating modes. Consequently, quasi-resonances should dominate energy transfer in large enough lattices, leading to the universal scaling law $T_c \propto g^{-2}$ in the thermodynamic limit.
	
For a fixed system size, the quasi-resonance connectivity strength decreases monotonically with $\Omega$, crossing below the exact resonance strength at a critical value $\Omega_c$. Below to this threshold, the exact resonance connectivity remains constant, establishing three distinct regimes: (i) the quasi-resonance-dominated regime exhibiting $T_c \propto g^{-2}$ scaling, (ii) the exact-resonance-dominated regime immediately below $\Omega_c$, where the identical scaling law transiently holds but ultimately terminates due to pairwise equalization, and (iii) the breakdown regime where higher-order processes emerge.  

 	\section{Validation of the pairwise equalization on the precise dynamics}
	
	\begin{figure*}[htbp] 
		\hspace*{-2cm}
		\centering
		\includegraphics{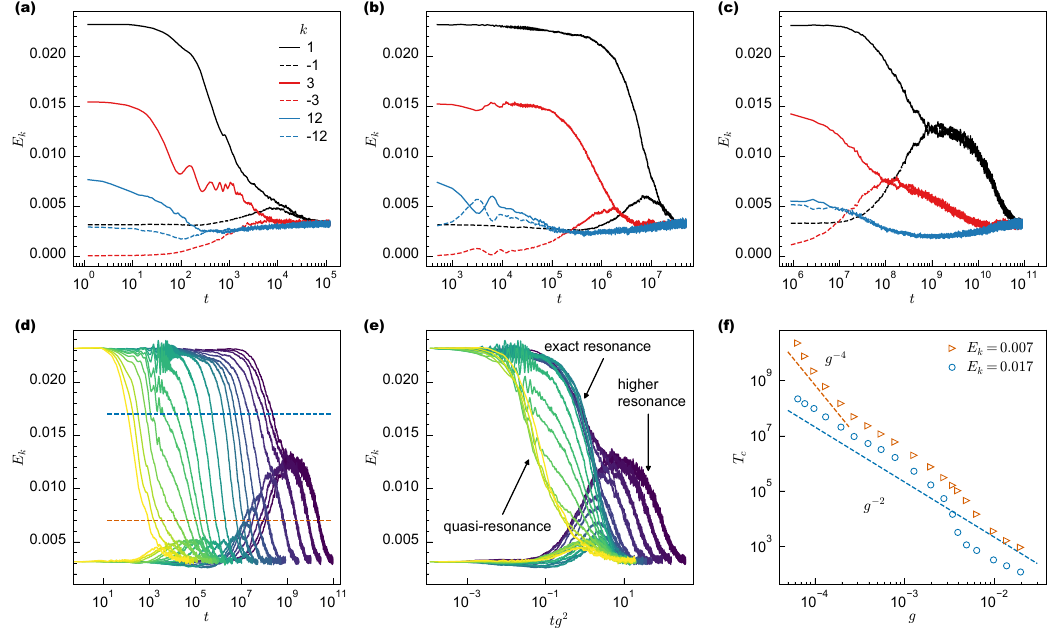}
		\caption{\label{fig:Ek-t-6p} Symmetry constraints in the precise dynamics of an $N=32$ FPUT-6 lattice with periodic boundary conditions. (a), (b), and (c) show the evolution of three symmetry mode pairs at \( g = 10^{-2},\ 5 \times 10^{-4},\ 6 \times 10^{-5} \), respectively. (d) shows the energy evolution of the \( k = \pm 1 \) modes over time for different nonlinearity strengths. From left to right, \( g \) gradually decreases from 0.02 to \( 6 \times 10^{-5} \). (e) shows the results after rescaling time by \( g^{-2} \). (f) shows the relaxation time \( T_c \), measured at the energy levels \( E_k = 0.017 \) and \( E_k = 0.007 \) as marked in (d), plotted as a function of \( g \). }
	\end{figure*}

Although the kinetic equations offer a theoretical framework for analyzing resonance dynamics, it deviates from the original lattice equations due to inherent random-phase and random-amplitude approximations. Crucially, this approach neglects quasi-resonant effects and higher-order nonlinearities. To verify the existence of symmetry recovery in actual lattice dynamics, we numerically solve the precise equations derived from Hamiltonian (Eq. 2), with a first focus on a 6-wave resonance set $(1,3,12,-1,-3,-12)$. To suppress stochastic fluctuations while preserving deterministic dynamics, an ensemble of realizations with random phases is applied (Methods). In this section, the lattice size is fixed at $N=32$ with periodic boundary conditions, and the energy of the modes is randomly assigned.

Figures~\ref{fig:Ek-t-6p}(a)-3(c) present the time evolution of mode energies, demonstrating that the pairwise equalization phenomenon persists in different nonlinearity strength magnitude. Meanwhile, there are several differences comparing with the the kinetic equation predictions. First, the full dynamics exhibits sustained energy decay after pairwise equalization, ultimately driving the system toward energy equipartition. Second, the pair equal-energy state preferentially initiates in higher-frequency mode pairs — in kinetic equation prediction all symmetric pairs recover symmetry nearly simultaneously(see Fig.~\ref{fig:jq}). Finally, the smaller the nonlinear intensity, the more similar the performance of pairwise equalization is to the kinetic equation prediction. In regions of strong nonlinearity, the pair of modes at $k=1$ and $k=-1$ do not strictly achieve energy equality until the energy is fully partitioned. In the case of weak nonlinearity, once energy equality is reached, it is maintained for a period of slow decay.

Figure~\ref{fig:Ek-t-6p}(d) displays the time evolution of the \(k = \pm 1\) mode pair energies across varying nonlinearity strengths \(g\), and Fig.~\ref{fig:Ek-t-6p}(e) provides the \(g^2\)-rescaled plot. These two plots show that the \(E_k\)'s separate into distinct clusters with \(T_c \propto g^{-2}\) scaling at both larger and smaller nonlinearity strengths. We observe that the relaxation timescale in the larger \(g\) regime maintains the \(g^{-2}\) scaling throughout the entire decay process and exhibits significantly faster decay. In the smaller \(g\) regime, the \(g^{-2}\) scaling is maintained over a certain range during the entire evolution, but in the extremely small \(g\) regime, this scaling holds in the early stage of evolution and begins to break down as the system approaches the pair equal-energy state. Compared to the larger \(g\) regime, the relaxation time is much longer in the latter cluster.

By employing two distinct energy thresholds (\(E_c = 0.017\) and \(E_c = 0.007\), marked in Fig.~\ref{fig:Ek-t-6p}(d)) to measure relaxation timescales, Fig.~\ref{fig:Ek-t-6p}(f) shows the relaxation time as a function of \(g\), respectively. Key observations are that, at \(E_c = 0.017\), the clean separation of \(g^{-2}\) scaling regimes directly maps to the two clusters in Fig.~\ref{fig:Ek-t-6p}(d) and 3(f). At \(E_c = 0.007\), while the large-\(g\) region maintains two segments of \(g^{-2}\) scaling, the small-\(g\) regime exhibits clear deviations.

Taking into account these results and Sec.~II, we can conclude that the scaling in the large-\(g\) regime is dominated by 6-wave quasi-resonances, while in the small-\(g\) regime, the scaling is governed by 6-wave exact resonances. In the latter case, the observed \(g^{-2}\) scaling throughout the full evolution process at larger \(g\) is likely due to the suppression of exact pairwise equalization under quasi-resonances and higher-order effects. At even smaller \(g\), higher-order nonlinear effects begin to dominate the relaxation, thus breaking the \(g^{-2}\) scaling, signaling the failure of exact 6-wave resonance. The analysis in Sec.~II already indicates that the strength of exact resonances decreases as the system size increases, suggesting that these exact resonance effects are only significant in finite, small-scale systems, and become negligible in sufficiently large systems.

	\section{Symmetry constraint effect in thermalization behavior}
	
	To demonstrate how symmetry constraint affects macroscopic properties, we systematically investigate the scaling laws of thermalization time $T_\text{eq}$ with respect to both nonlinearity strength $g$ and system size $N$. Following the established protocol in references~\cite{Onorato:2015,Pistone:2018}, we compute entropy $s(t)$ and define $T_\text{eq}$ as the time when $\langle s(t) \rangle/(N-1) = 0.1$ is reached. 
	
	\begin{figure}[htb] 
		\centering
		\includegraphics{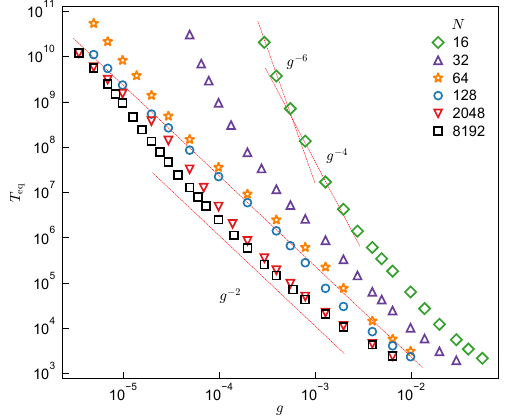}  	
		\caption{Thermalization time of FPUT-$6$ lattices as a function of nonlinearity strength. Dashed lines indicate typical slopes for reference.}\label{fig:T} 	
	\end{figure}
	
Figure 4 presents the key findings for our FPUT-6 lattice model. For smaller systems ($N = 16,32,64$), the characteristic $g^{-2}$ scaling persists across a broad range of $g$, but significant deviations emerge in the weak nonlinearity regime. Most notably, the $N=16$ system exhibits successive transitions between distinct scaling laws: from the initial $T_\text{eq} \propto g^{-2}$ (characterizing the 6-wave resonances) to $T_\text{eq} \propto g^{-4}$ (characterizing the 10-wave) and eventually $T_\text{eq} \propto g^{-6}$ (characterizing the 14-wave) as $g$ decreases (Supplementary Section 4). In intermediate systems ($N=128$), a single apparent $g^{-2}$ scaling persists across the entire regime. In sufficiently large systems ($N = 2048,8192$), three distinct regimes emerge: in the larger $g$ regime, the scaling approaches $g^{-1}$; in the intermediate $g$ regime, it approaches a fast relaxation with $g^{-2}$ scaling; and in the smaller $g$ regime, it transitions to a slow relaxation with $g^{-2}$ scaling.

These main results can be understood based on the dependence of exact and quasi-resonances on system size and nonlinearity strength, as revealed in Fig. 2. At small sizes, the exact resonance connectivity strength is relatively large, leading to the occurrence of pairwise equalization, which terminates the exact 6-wave resonance. As \( g \) decreases, the system gradually transitions to a relaxation process dominated by higher-order resonances, deviating from the \( g^{-2} \) scaling. In intermediate-sized systems, the 6-wave quasi-resonances maintain the 6-wave resonance scaling across a broad parameter range. Since the exact resonance connectivity strength remains significant in this regime, the 6-wave quasi-resonance dominated region merges with the exact 6-wave resonance dominated region, leading to the failure to distinguish which \( g^{-2} \) scaling corresponds to the 6-wave quasi-resonance and which corresponds to the exact 6-wave resonance. For sufficiently large systems, the exact resonance connectivity strength becomes very low. In this case, higher-order effects surpass the exact 6-wave resonance effects after the failure of the 6-wave quasi-resonances, resulting in a sudden increase in thermalization time. Once \( g \) decreases further and the higher-order resonance strength falls below that of exact 6-wave resonance, the \( g^{-2} \) scaling reappears. Further reduction of \( g \) or using a more small $\langle s(t) \rangle/(N-1) $ threshold will cause this scaling law to fail due to the pairwise equalization effect, meaning that in finite-sized systems, the exact 6-wave resonance will eventually break down.
	
The observed deviation from the \( g^{-2} \) scaling in bigger $g$ regime likely stems from Chirikov resonance mechanisms. These resonances occur when nonlinear frequency broadening bridges the gap between adjacent modal frequencies, enabling enhanced pairwise energy transfer and accelerated thermalization~\cite{Wang:2024}. This phenomenon, well-documented across various nonlinear lattices~\cite{Lvov:2018,Wang:2024}, exhibits characteristic $N$-dependence: the decreasing frequency spacing in larger systems facilitates Chirikov overlap, while smaller systems remain unaffected. Consequently, thermalization in this regime represents a hybrid process combining Chirikov-type pairwise resonances with 6-wave quasi-resonant interactions. 
    
	\section{Universality}
	
	\begin{figure*}[htbp] 
		\hspace*{-0.8cm}
		\centering
		\includegraphics{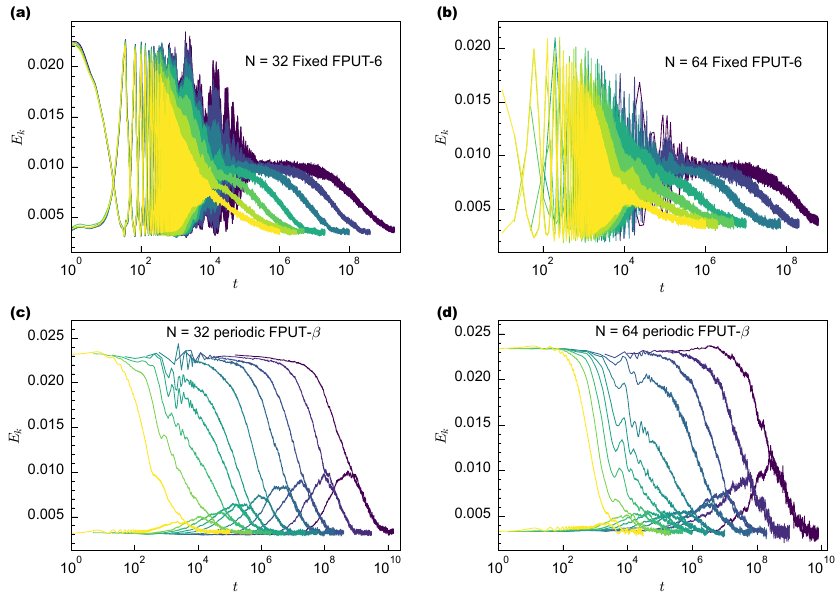}
\caption{\label{fig:5} 
Universality of symmetry constraints. (a) and (b) show the time evolution of the energy for the \( k = \pm 1 \) modes in the fixed-boundary FPUT-6 model, while (c) and (d) show the corresponding results for the periodic-boundary FPUT-$\beta$ model, under varying nonlinearity strengths. From left to right, \( g \) decreases from \( 3 \times 10^{-3} \) to \( 10^{-4} \) in (a) and (b), and from \( 0.13 \) to \( 0.004 \) in (c) and (d).}
\end{figure*}

A key step to confirm that the symmetry constraint is generally effective is to demonstrate that this phenomenon also exists under fixed boundary conditions. In principle, the symmetry constraint is determined by the symmetries of the kinetic equations and is independent of the boundary conditions. However, under fixed boundary conditions, $k$ and $-k$ traveling wave pairs can only exist during transient processes. 

To show that it is the pairwise equalization between $k$ and $-k$ traveling waves that breaks down the exact resonances and leads to the formation of standing waves, we impose traveling waves with different amplitudes for the $-k$ and $k$ modes within the lattice length to drive the system away from the standing wave solution, and then track the evolution process. The evolution of the \( k = \pm 1 \) mode pair across varying nonlinearity strengths \( g \) is shown in Fig.~\ref{fig:5}(a) and 5(b) for $N=32$ and $N=64$, respectively. During the evolution, we project the particle coordinates and momenta onto two traveling waves of $k=1$ and $k=-1$ propagating in opposite directions at specific eigenmode frequencies.

Due to the repeated reflections of traveling waves between fixed boundaries, large-amplitude energy oscillations occur, making the mutual energy exchange between the \( k = \pm 1 \) modes difficult to observe. However, as time evolves, we observe clear pairwise equalization behavior. The time required to reach the state \( E_k = E_{-k} \) appears to be much shorter than in systems with periodic boundary conditions, which can be attributed to the enhanced interactions caused by frequent boundary reflections. As the system size increases, this equalization time grows rapidly. It is expected that, for sufficiently large systems, the pairwise equalization dynamics will become indistinguishable from those observed under periodic boundary conditions.

	Similarly, the symmetry constraint is not restricted to any specific model or limited to the 6-wave resonance scenario, as the symmetry properties of the kinetic equations are fundamentally general. We further demonstrate the universality of this phenomenon by examining the FPUT-$\beta$ model—one of the most extensively studied systems in FPUT research. The Hamiltonian of this model is given by:  
	\[
	H = \sum_{j} \left[ \frac{p_j^2}{2} + \frac{1}{2}(q_j - q_{j+1})^2 + \frac{1}{4}\beta (q_j - q_{j+1})^4 \right],
	\]  
where the nonlinearity strength is characterized by \( g = \beta \epsilon \).  
	
	Illustrative examples are shown in Fig.~\ref{fig:5}(c) and Fig.~5(d), where we track the evolution of mode pairs with \( k = \pm 1 \) across varying nonlinearity strengths \( g \) in lattices of size \( N = 32 \) and \( N = 64 \), respectively. The results exhibit striking similarity to those in Fig.~3(d), with symmetry recovery persisting uniformly across the entire parameter range. Note that the lowest-order nonlinear term in this system is quartic, and therefore the primary resonance corresponds to a 4-wave interaction.
	
	\section{Conclusion and Discussion}

In finite systems, exact resonances are terminated during the evolution due to symmetry constraints, which confine the energy to specific symmetric subsets of modes. For sufficiently large systems, however, the connectivity strength of the exact resonance network decreases with system size, rendering its effect negligible. Therefore, the textbook-level exact multi-wave resonance conditions are physically incomplete. 

In contrast, quasi-resonances constitute the fundamental mechanism governing energy diffusion or transfer among modes in lattices. Since the connectivity strength of the quasi-resonance network increases with system size, arbitrarily small nonlinear interactions can lead to thermalization in the thermodynamic limit, following a universal thermalization time scaling law. 

On the other hand, as the nonlinearity strength decreases, the connectivity of the quasi-resonance network also diminishes, leading to the breakdown of $n$-wave resonances and a transition to an energy transfer network dominated by $(n+1)$-wave resonances. This results in a finite-size threshold for nonlinearity, below which thermalization fails as the thermalization time diverges. This threshold behavior confirms the stretched-exponential phenomenology observed in Nekhoroshev-type systems~\cite{Nekhoroshev:1977,Benettin:1984,Berchialla:2004}. The symmetry constraint effect is not restricted to specific models and is expected to universally exist in generic lattice systems.

For the FPUT problem, this implies that if the nonlinearity strength is sufficiently weak, finite-size models will indeed exhibit persistent FPUT recurrences and fail to reach thermal equilibrium. Only in the thermodynamic limit can FPUT-type lattices fulfill the equipartition theorem. This conclusion not only provides a comprehensive answer to the FPUT paradox, but also establishes a theoretical foundation for understanding whether, and under what conditions, nonequilibrium systems can evolve toward equilibrium. It further suggests the necessity of reexamining and reformulating multi-phonon resonance conditions within the framework of quantum mechanics.

	\bibliography{ref}
	
	
	\section*{Methods}
	\subsubsection*{Derivation of the kinetic equations}
	
	The rescaled Hamiltonian of the FPUT-6 lattice takes the form 
	\begin{eqnarray}\label{eq:Ht-6}
		\begin{split}
			\tilde{H} =& \sum_{j} \left[ \frac{\tilde{p}_j^2}{2} +  \frac{1}{2}(\tilde{q}_{j+1} - \tilde{q}_j)^2  + \frac{1}{6} g (\tilde{q}_j -\tilde{q}_{j+1})^6 \right]\\
			=& H_0(\tilde{q}_j,\tilde{p}_j) + \sum_{j} \frac{1}{6} g (\tilde{q}_j -\tilde{q}_{j+1})^6.
		\end{split}
	\end{eqnarray}
	The Hamiltonian of the harmonic part $H_0(\tilde{q}_j,\tilde{p}_j)$ can be diagonalized to obtain the system's eigenmodes $u^k_j$ and eigenfrequencies $\omega_k$. By projecting the displacements and momenta onto these eigenmodes, one can obtain the normal coordinate $Q_k = \sum_j \tilde{q}_j u^k_j$ and the normal momentum $P_k = \dot{Q}_k = \sum_j \tilde{p}_j u^k_j$. Introducing the complex normal variable $a_k = \frac{1}{\sqrt{2\omega_k}}(\omega_k Q_k + i P_k)$, the Hamiltonian can be rewritten as
	\begin{eqnarray}\label{eq:Hak-6}
		\begin{split}
			\tilde{H} = \sum_{k} \omega_k a_k a^*_k + \frac{g}{6}  \sum_{k_1,..,k_6} A_{k_1k_2k_3k_4k_5k_6} \prod_{s=1}^6 (a_{k_s}+a^*_{k_s}),
		\end{split}
	\end{eqnarray}
	where $A_{k_1k_2k_3k_4k_5k_6} = \frac{1}{8 \sqrt{\omega_{k_1} \omega_{k_2} \omega_{k_3} \omega_{k_4}\omega_{k_5}\omega_{k_6}}} \sum_j \prod_{s=1}^6 \left(u^{k_s}_j- u^{k_s}_{j+1}\right)$. For a homogeneous lattice, one can derive the 6-wave resonance conditions for the wave vectors from  $A_{k_1k_2k_3k_4k_5k_6}$. For example, under periodic boundary conditions,  the eigenmodes of the lattice are given by $u^{k_s}_j=\sqrt{\frac{2}{N}}\cos\left(\frac{2k_s\pi j}{N}\right)=\frac{1}{\sqrt{2N}}\left(e^{\frac{i2k_s\pi j}{N}}+e^{\frac{-i2k_s\pi j}{N}}\right)$. Inserting this expression into the definition of $A_{k_1k_2k_3k_4k_5k_6}$, we obtain:
		\begin{eqnarray}\label{eq:A123456}
		\begin{split}
			&A_{k_1k_2k_3k_4k_5k_6}\\
			=& \frac{1}{64 N^3 \sqrt{\omega_{k_1} \omega_{k_2} \omega_{k_3} \omega_{k_4}\omega_{k_5}\omega_{k_6}}} \sum_j \prod_{s=1}^6 \left(e^{\frac{i2k_s\pi j}{N}} +e^{-\frac{i2k_s\pi j}{N}}  -  e^{\frac{i2k_s\pi (j-1)}{N}} - e^{-\frac{i2k_s\pi (j-1)}{N}} \right)\\
			=&\frac{1}{64N^3 \sqrt{\omega_{k_1} \omega_{k_2} \omega_{k_3} 	\omega_{k_4}\omega_{k_5}\omega_{k_6}}} \sum_j \prod_{s=1}^6 \left(e^{\frac{i2k_s\pi }{N}}  -e^{\frac{-i2k_s\pi}{N}} \right)\left(e^{\frac{ik_s\pi (2j-1)}{N}} - e^{\frac{-ik_s\pi (2j-1)}{N}}  \right)\\
			=&\pm \frac{1}{64N^3 \sqrt{\omega_{k_1} \omega_{k_2} \omega_{k_3} 	\omega_{k_4}\omega_{k_5}\omega_{k_6}}} \sum_j \prod_{s=1}^6 \omega_{k_s} \left(e^{\frac{ik_s\pi (2j-1)}{N}} - e^{\frac{-ik_s\pi (2j-1)}{N}}\right)\\
			=&\pm\frac{\sqrt{\omega_{k_1} \omega_{k_2} \omega_{k_3} 	\omega_{k_4} \omega_{k_5}\omega_{k_6}}}{64 N^3}  \sum_j \cos\left[\frac{(2j-1)(\pm k_1 \pm k_2 \pm k_3 \pm k_4 \pm k_5 \pm k_6)\pi }{N}\right]\\
			=& \frac{\sqrt{\omega_{k_1} \omega_{k_2} \omega_{k_3} 	\omega_{k_4} \omega_{k_5}\omega_{k_6}}}{64 N^2} \left(\pm \delta_{k_1}^{k_2k_3k_4k_5k_6} \pm \delta_{k_1k_2}^{k_3k_4k_5k_6} \pm  \delta_{k_1k_2k_3}^{k_4k_5k_6} \right. \left. \pm \delta_{k_1k_2k_3k_4}^{k_5k_6} \pm  \delta_{k_1k_2k_3k_4k_5}^{k_5k_6}\pm \delta_{k_1k_2k_3k_4k_5k_6}^{}\right).
		\end{split}
	\end{eqnarray}
    Here, $\delta^{k_1k_2k_3}_{k_4k_5k_6}=\delta(k_1+k_2+k_3-k_4-k_5-k_6)$, other terms in the brackets
	can be expressed similarly. From this, the 6-wave resonance conditions for the wave vectors can be obtained as
	\begin{equation}\label{eq:6bo-k}
		k_1 \pm k_2  \pm k_3 \pm k_4 \pm k_5 \pm k_6 \ \ \text{mod} \ \ N\  = \ 0.
	\end{equation}
	
	The dynamical equation can be expressed as
	\begin{eqnarray}\label{eq:ia6k}
		i\dot{a}_{k_1} = \frac{\partial H}{\partial a^*_{k_1}} = \omega_{k_1}a_{k_1} + g \sum_{k_2,...,k6} A_{k_1k_2k_3k_4k_5k_6} \prod_{s=2}^6 (a_{k_s}+a^*_{k_s}).
	\end{eqnarray}
	Multiplying Eq.~(\ref{eq:ia6k}) with $a_{k_1}^*$ and taking the ensemble average, we derive the kinetic equation for 6-wave resonance as
	\begin{eqnarray}\label{eq:H6-ke0}
		\begin{split}
			\dot{D}_{k_1}=&\langle \dot{a}_{k_1}a^*_{k_1}+\dot{a}^*_{k_1}a_{k_1}\rangle=2 {\rm Im} \left[\langle\dot{a}_{k_1}a^*_{k_1}\rangle \right] \\
			= &2g  {\rm Im}\left[\sum_{k_2,...,k_6} A_{k_1k_2k_3k_4k_5k_6}  \langle a^*_{k_1} \prod_{s=2}^6 (a_{k_s}+a^*_{k_s})\rangle  \right]\\
			= &2g  {\rm Im}\left[\sum_{k_2,...,k_6} A_{k_1k_2k_3k_4k_5k_6}  \langle a^*_{k_1} a_{k_2} a_{k_3}a_{k_4}a_{k_5}a_{k_6}+C_5^1a^*_{k_1} a^*_{k_2} a_{k_3}a_{k_4}a_{k_5}a_{k_6}  \right.\\
			&+C_5^2a^*_{k_1} a^*_{k_2} a^*_{k_3}a_{k_4}a_{k_5}a_{k_6} +C_5^3a^*_{k_1} a^*_{k_2} a^*_{k_3}a^*_{k_4}a_{k_5}a_{k_6} \\
			&+ C_5^4a^*_{k_1} a^*_{k_2} a^*_{k_3}a^*_{k_4}a^*_{k_5}a_{k_6}+ C_5^5a^*_{k_1} a^*_{k_2} a^*_{k_3}a^*_{k_4}a^*_{k_5}a^*_{k_6} \rangle \Bigg].
		\end{split}
	\end{eqnarray}
	Here, $\text{Im}()$ denotes taking the imaginary part of the complex number inside the parentheses. Under the stochastic phase and stochastic amplitude assumptions~\cite{Nazarenko:2011}, except for the zeroth-order term of the correlation function $\langle a^*_{k} a^*_{l} a^*_{m}a_{n}a_{o}a_{p} \rangle  = D_k D_l D_m (\delta^k_n \delta^l_o \delta^m_p + ...) $ being real, all other terms are zero. This indicates that if only first-order perturbations are considered, then $\dot{D}_{k_1}=0$. Therefore, we need to calculate $\dot{D}_{k_1}$ up to the second-order perturbation. For this, it is necessary to know the sixth-order correlation functions to first order.  For the 3-3 process, we can take the partial derivative with respect to time of $\langle a^*_{k} a^*_{l} a^*_{m}a_{n}a_{o}a_{p} \rangle$  and substitute Eq.~(\ref{eq:ia6k}) into it, then move the terms containing the frequency to the left side of the equation, and combine like terms, thereby obtaining: 
	\begin{eqnarray}\label{eq:iaaa-6}
		\begin{split}
			&\left[i\frac{\partial}{\partial t} + \left(\omega_{k_1} + \omega_{k_2} + \omega_{k_3} - \omega_{k_4}-\omega_{k_5}-\omega_{k_6}\right) \right]\langle a^*_{k_1} a^*_{k_2} a^*_{k_3}a_{k_4} a_{k_5}a_{k_6}\rangle \\
			= & i\langle \dot{a}^*_{k_1} a^*_{k_2} a^*_{k_3}a_{k_4} a_{k_5}a_{k_6}\rangle +
			i\langle a^*_{k_1} \dot{a}^*_{k_2} a^*_{k_3}a_{k_4} a_{k_5}a_{k_6}\rangle +
			i\langle a^*_{k_1} a^*_{k_2} \dot{a}^*_{k_3}a_{k_4} a_{k_5}a_{k_6}\rangle  \\
			&+ i\langle a^*_{k_1} a^*_{k_2} a^*_{k_3}\dot{a}_{k_4} a_{k_5}a_{k_6}\rangle + i\langle a^*_{k_1} a^*_{k_2} a^*_{k_3}a_{k_4} \dot{a}_{k_5}a_{k_6}\rangle+ i\langle a^*_{k_1} a^*_{k_2} a^*_{k_3}a_{k_4} a_{k_5}\dot{a}_{k_6}\rangle\\
			& + \left(\omega_{k_1} + \omega_{k_2} + \omega_{k_3} - \omega_{k_4}-\omega_{k_5}-\omega_{k_6}\right)\langle a^*_{k_1} a^*_{k_2} a^*_{k_3}a_{k_4} a_{k_5}a_{k_6}\rangle  \\
			=&  - \langle g \sum_{k_2,k_3,k_4,k_5,k_6} A_{k_1k_2k_3k_4k_5k_6} a^*_{k_2} a^*_{k_3}a_{k_4} a_{k_5}a_{k_6} \prod_{s=2,3,4,5,6} (a^*_{l_s}+a_{l_s})\rangle - ... \\
			=& -120 g  A_{k_1k_2k_3k_4k_5k_6} \langle a^*_{k_2} a^*_{k_3}a_{k_4} a_{k_5}a_{k_6}a_{k_2} a_{k_3}a^*_{k_4} a^*_{k_5}a^*_{k_6} \rangle  - ... \\
			=& - 120g  A_{k_1k_2k_3k_4k_5k_6} \left(D_{k_2} D_{k_3} D_{k_4} D_{k_5} D_{k_6} +...-D_{k_1} D_{k_2} D_{k_3} D_{k_4} D_{k_5}\right) \\
			=&  120g  A_{k_1k_2k_3k_4k_5k_6}  D_{k_1} D_{k_2} D_{k_3} D_{k_4} D_{k_5} D_{k_6} \left(\frac{1}{D_{k_4}} + \frac{1}{D_{k_5}}+ \frac{1}{D_{k_6}} -\frac{1}{D_{k_1}}-\frac{1}{D_{k_2}} -\frac{1}{D_{k_3}}\right) \\
			=&  g A_{k_1k_2k_3k_4k_5k_6} C^{k_4k_5k_6}_{k_1k_2k_3}, 
		\end{split}
	\end{eqnarray}
	where $C^{k_4k_5k_6}_{k_1k_2k_3} =  120D_{k_1} D_{k_2} D_{k_3} D_{k_4} D_{k_5} D_{k_6} \left(\frac{1}{D_{k_4}} + \frac{1}{D_{k_5}}+ \frac{1}{D_{k_6}} -\frac{1}{D_{k_1}}-\frac{1}{D_{k_2}} -\frac{1}{D_{k_3}}\right)$. 
	Using $\lim_{\delta \to 0} 	{\rm Im}\left( \frac{1}{i\delta + \omega} \right) = -\pi \delta(\omega)$, Eq.~(\ref{eq:iaaa-6}) can be written as
	\begin{eqnarray}\label{eq:ia3-3}
		\begin{split}
			{\rm Im} \left(\langle a^*_{k_1} a^*_{k_2} a^*_{k_3}a_{k_4} a_{k_5}a_{k_6}\rangle \right) = &{\rm Im} \left( \frac{g A_{k_1k_2k_3k_4k_5k_6} C^{k_4k_5k_6}_{k_1k_2k_3}}{\left[i\frac{\partial}{\partial t} + \left(\omega_{k_1} + \omega_{k_2} + \omega_{k_3} - \omega_{k_4}-\omega_{k_5}-\omega_{k_6}\right)  \right]} \right) \\
			=& -\pi g A_{k_1k_2k_3k_4k_5k_6} C^{k_4k_5k_6}_{k_1k_2k_3} \omega_{k_1k_2k_3}^{k_4k_5k_6} 
		\end{split}
	\end{eqnarray}
	where $\omega_{k_1k_2k_3}^{k_4k_5k_6}= \delta ( \omega_{k_4} + \omega_{k_5} + \omega_{k_6} - \omega_{k_1} - \omega_{k_2} - \omega_{k_3})$.
	
	The additional terms in Eq.~(\ref{eq:H6-ke0}) can be obtained using the similar operation. For example, \begin{eqnarray}\label{eq:ia2-4}
		\begin{split}
			&{\rm Im} \left(\langle a^*_{k_1} a^*_{k_2} a_{k_3}a_{k_4} a_{k_5}a_{k_6}\rangle \right)\\ 
			=&{\rm Im} \left( \frac{g A_{k_1k_2k_3k_4k_5k_6} C^{k_3k_4k_5k_6}_{k_1k_2}}{\left[i\frac{\partial}{\partial t} + \left(\omega_{k_1} + \omega_{k_2} - \omega_{k_3} - \omega_{k_4}-\omega_{k_5}-\omega_{k_6}\right)  \right]} \right) \\
			=& -\pi g A_{k_1k_2k_3k_4k_5k_6} C^{k_3k_4k_5k_6}_{k_1k_2} \omega_{k_1k_2}^{k_3k_4k_5k_6}
		\end{split}
	\end{eqnarray}
	for the 2-4 process, where $C^{k_3k_4k_5k_6}_{k_1k_2} =  120 D_{k_1} D_{k_2} D_{k_3} D_{k_4} D_{k_5} D_{k_6} \left(\frac{1}{D_{k_3}}+\frac{1}{D_{k_4}} + \frac{1}{D_{k_5}} \right.$  $\left.+ \frac{1}{D_{k_6}} -\frac{1}{D_{k_1}}-\frac{1}{D_{k_2}} \right)$, $\omega_{k_1k_2}^{k_3k_4k_5k_6}= \delta ( \omega_{k_3}+ \omega_{k_4} + \omega_{k_5} + \omega_{k_6} - \omega_{k_1} - \omega_{k_2})$. And for the 4-2 proces, \begin{eqnarray}\label{eq:ia4-2}
		\begin{split}
			&{\rm Im} \left(\langle a^*_{k_1} a^*_{k_2} a^*_{k_3}a^*_{k_4} a_{k_5}a_{k_6}\rangle \right) \\
			=&{\rm Im} \left( \frac{g A_{k_1k_2k_3k_4k_5k_6}C^{k_5k_6}_{k_1k_2k_3k_4}}{\left[i\frac{\partial}{\partial t} + \left(\omega_{k_1} + \omega_{k_2} + \omega_{k_3} + \omega_{k_4}-\omega_{k_5}-\omega_{k_6}\right)  \right]} \right) \\
			=& -\pi g A_{k_1k_2k_3k_4k_5k_6} C^{k_5k_6}_{k_1k_2k_3k_4} \omega_{k_1k_2k_3k_4}^{k_5k_6},
		\end{split}
	\end{eqnarray}
	where $C^{k_5k_6}_{k_1k_2k_3k_4}=  120 D_{k_1} D_{k_2} D_{k_3} D_{k_4} D_{k_5} D_{k_6}\left(\frac{1}{D_{k_5}}+ \frac{1}{D_{k_6}} -\frac{1}{D_{k_1}}-\frac{1}{D_{k_2}}-\frac{1}{D_{k_3}}-\frac{1}{D_{k_4}}  \right)$, $\omega_{k_1k_2k_3k_4}^{k_5k_6}= \delta (\omega_{k_5} + \omega_{k_6} - \omega_{k_1} - \omega_{k_2} - \omega_{k_3} - \omega_{k_4})$.
	By incorporating all expressions of the sixth-order correlation functions into Eq.~(\ref{eq:H6-ke0}), the 6-wave kinetic equation can be rewritten as \begin{eqnarray}\label{eq:H6keq-0}
		\begin{split}
			\dot{D}_{k_1}= & -2g^2 \pi \sum_{k_2,...,k_6} A^2_{k_1k_2k_3k_4k_5k_6}(C^{k_2k_3k_4k_5k_6}_{k_1}\omega^{k_2k_3k_4k_5k_6}_{k_1}  \\
			& + 5C_{k_1k_2}^{k_3k_4k_5k_6}\omega_{k_1k_2}^{k_3k_4k_5k_6} +  10C^{k_4k_5k_6}_{k_1k_2k_3}\omega_{k_1k_2k_3}^{k_4k_5k_6} +10C_{k_1k_2k_3k_4}^{k_5k_6}\omega_{k_1k_2k_3k_4}^{k_5k_6} \\
			&+ 5C_{k_1k_2k_3k_4k_5}^{k_6}\omega^{k_6}_{k_1k_2k_3k_4k_5}+C_{k_1k_2k_3k_4k_5k_6}\omega_{k_1k_2k_3k_4k_5k_6}).\\
		\end{split}
	\end{eqnarray}
	By decomposing terms  $C^{k_2k_3k_4k_5k_6}_{k_1},C_{k_1k_2}^{k_3k_4k_5k_6},C^{k_4k_5k_6}_{k_1k_2k_3},C_{k_1k_2k_3k_4}^{k_5k_6},C_{k_1k_2k_3k_4k_5}^{k_6},C_{k_1k_2k_3k_4k_5k_6}$ into parts related and unrelated to $D_{k_1}$, Eq.~(\ref{eq:H6keq-0}) can be further simplified to
	\begin{eqnarray}\label{eq:6keq}
		\begin{split}
			\dot{D}_{k_1} = \eta_{1} -  \gamma_{1}  D_{k_1}.
		\end{split}
	\end{eqnarray}
	Both $\eta_{1}$ and $\gamma_{1}$ are independent of $D_{k_1}$ and are both proportional to $g^2$. From frequency-related terms in Eq.~(\ref{eq:H6keq}), such as $\omega_{k_1k_2}^{k_3k_4k_5k_6}, \omega_{k_1k_2k_3}^{k_4k_5k_6}, \omega_{k_1k_2k_3k_4}^{k_5k_6}$, the frequency conditions for 6-wave resonance can be derived, i.e., 
	\begin{equation}\label{eq:6bo-w}
		\omega_{k_1} \pm \omega_{k_2} \pm \omega_{k_3} \pm\omega_{k_4} \pm \omega_{k_5} \pm\omega_{k_6}  = 0.
	\end{equation}

	\subsubsection*{Evolution of the kinetic equations}
	
	At the initial moment of the simulation, we set the energy of the $k = 0$ mode to $E_k = 0$, and the energies of the remaining modes, $E_k$, are uniformly and randomly selected within the range $[0, 0.006]$, maintaining an energy density of $\epsilon = \sum E_k / (N-1) = 0.003$. The reason for using $N-1$ instead of $N$ can be seen from the kinetic equations: the eigenfrequency of the $k = 0$ mode, $\omega_{k_1} = 0$, directly leads to $A_{k k_2 k_3 k_4 k_5 k_6} = 0$, i.e., $\dot{D}_k = 0$. Therefore, the $k = 0$ mode does not participate in the evolution of the kinetic equations, and the actual number of independent degrees of freedom is $N-1$. The nonlinearity strength $g$ is adjusted by varying $b$.

	The wave action spectral density can be obtained through the relation of $D_{k} = E_{k} / \omega_{k}$. Next, by substituting the values of $D_{k}$ into the kinetic equations and iterating over $k_1, \dots, k_6$ that satisfy the nontrivial exact resonance solutions (such as 3-3 symmetry solutions, 3-3 quasi-symmetry solutions, etc.), the value of $\dot{D}_{k_1}$ for each $k_1$ mode can be obtained. By performing an Euler integration on $\dot{D}_{k_1}(t_0)$ at time $t_0$, i.e., $D_{k_1}(t_1) = D_{k_1}(t_0) + \dot{D}_{k_1}(t_0) dt$, we can estimate the value of $D_{k_1}(t_1)$ at time $t_1$. By substituting $D_{k_1}(t_1)$ back into the kinetic equation, we obtain the value of $\dot{D}_{k_1}(t_1)$, which can then be integrated to yield $D_{k_1}(t_2)$. Repeating this process iteratively yields the time evolution of $D_{k_1}$, or equivalently, the energy $E_{k_1}$. This procedure can simulate the energy diffusion behavior between coupled modes under the exact resonance solutions. In the numerical integration process, various values of the time step $dt$ were tested. The simulation started with a relatively large time step, which was gradually reduced until the chosen time step no longer affected the results.

	\subsubsection*{Evolution of the dynamical equations}
	
	At the initial moment, the positions and momenta of the atoms in the lattice are assigned in the following way:
	\begin{eqnarray}
		\begin{split}
			q_j=& \sum_k  \frac{\sqrt{E_k}}{\omega_k} \sqrt{\frac{2}{N}} \cos(\frac{2k\pi j}{N}+\phi_k^{t_0}),\\
			p_j =& \sum_k \sqrt{E_k} \sqrt{\frac{2}{N}}  \sin(\frac{2k\pi j}{N}+ \phi^{t_0}_k).
		\end{split}	
	\end{eqnarray}
	Each mode is assigned a noise-level energy randomly. Then, we select a specific 6-wave resonance set, such as $(k_1, k_2, k_3, k_4, k_5, k_6) = (1, 3, 12, -1, -3, -12)$, and give relatively large energies to the $k_1, k_2, k_3$ modes. However, we ensure that the energy density is fixed, i.e., $\epsilon = 0.003$. The nonlinearity strength $g$ is similarly adjusted by varying $b$ for FPUT-6 and $\beta$ for FPUT-$\beta$. Then, the original dynamical equations of the system are integrated using the eighth-order Yoshida method\cite{Yoshida:1990}, and the positions and momenta of the atoms over time can be obtained. The typical integration time step is set to $\Delta t = 0.1$.
	
To distinguish the energies of the left- and right-traveling waves, we perform Fourier transforms on $q_j$ and $p_j$, and expand the sine and cosine functions into exponential forms using Euler’s formulas, $\cos(\theta) = \frac{e^{i \theta} + e^{-i \theta}}{2}$ and $\sin(\theta) = \frac{e^{i \theta} - e^{-i \theta}}{2i}$, respectively. After simplification, we obtain: 
	\begin{eqnarray}
		\begin{split}	
			\tilde{Q}_k =&\sum_{j=0}^{N-1} q_j e^{-i 2 \pi k j / N}   \\
			=&\frac{1}{\omega_k } \sqrt{\frac{E_kN}{2}} e^{i \phi_k^t} +\frac{1}{\omega_k} \sqrt{\frac{E_{-k}N}{2}} e^{-i \phi_{-k}^t}, \\
			\tilde{P}_k =&\sum_{j=0}^{N-1} p_j e^{-i 2 \pi k j / N}   \\
			=& -i\left(\sqrt{\frac{E_kN}{2}} e^{i \phi_k^t} - \sqrt{\frac{E_{-k}N}{2}} e^{-i \phi_{-k}^t} \right).
		\end{split}
	\end{eqnarray}	
	By combining the two expressions above, we obtain the energies of the right- and left-traveling waves, respectively:
	\begin{eqnarray}
		E_k = \frac{\left|\omega_k \tilde{Q}_k+i\tilde{P}_k\right|^2}{2N}, \  E_{-k} = \frac{\left|\omega_k \tilde{Q}_k - i\tilde{P}_k\right|^2}{2N}.
	\end{eqnarray}	
	
	Due to the large fluctuations in the precise dynamical system, we adopt an ensemble averaging strategy to demonstrate the trend of energy changes in these modes: For a fixed nonlinear parameter, we evolve 1024 sets of initial conditions and compute the ensemble average of $E_k$. Each set of initial energy distributions is the same, but the initial phases are uniformly distributed over the interval $[0, 2\pi]$. 
	
	\subsubsection*{The calculation of the indicator entropy}
	
	The indicator entropy $s(t)$ is defined as $s(t) = \sum_k f_k \ln(f_k)$~\cite{Onorato:2015}, where $f_k = E_k/\langle E_k \rangle$ and $\langle E_k \rangle = \sum E_k/(N-1)$. The use of $N - 1$ instead of $N$ is also due to the fact that the $k = 0$ mode does not participate in the thermalization process. As $E_k$ approaches the equilibrium value $\langle E_k \rangle$, $s(t)$ approaches 0. The degree of energy equipartition is thus measured by $\langle s(t) \rangle/(N-1)$. Here, the angle brackets represent ensemble averaging, meaning that $s(t)$ is averaged over the simulation results from different initial energy configurations (with random phases and amplitudes, while $g$ is held constant).

	
	

	
	
\end{document}